# Tunable Electronic Structure and Surface States in Rare Earth Mono-Bismuthides with Partially Filled *f* Shell

Peng Li[1#], Zhongzheng Wu[1#], Fan Wu[1#], Chao Cao[2*], Chunyu Guo[1], Yi Wu[1], Yi Liu[3], Zhe Sun[3], Cheng-Maw Cheng[4], Deng-Sung Lin[5], Frank Steglich[1], Huiqiu Yuan[1*], Tai-Chang Chiang[6] and Yang Liu[1*]

[1]Center for Correlated Matter and Department of Physics, Zhejiang University, Hangzhou, China

[2]Department of Physics, Hangzhou Normal University, Hangzhou, China

[3]National Synchrotron Radiation Laboratory, University of Science and Technology of China, Hefei, China

[4]National Synchrotron Radiation Research Center, Hsinchu, Taiwan

[5]Department of Physics, National Tsinghua University, Hsinchu, Taiwan

[6]Department of Physics and Frederick Seitz Material Research Lab, University of Illinois at Urbana-Champaign, Urbana, USA

#these authors contribute equally to the work

*Corresponding author: yangliuphys@zju.edu.cn, hqyuan@zju.edu.cn, ccao@hznu.edu.cn

# Abstract

Here we report the evolution of bulk band structure and surface states in rare earth mono-bismuthides with partially filled $f$ shell. Utilizing synchrotron-based photoemission spectroscopy, we determined the three-dimensional bulk band structure and identified the bulk band inversions near the $X$ points, which, according to the topological theory, could give rise to nontrivial band topology with odd number of gapless topological surface states. Near the surface $\overline{\Gamma}$ point, no clear evidence for predicted gapless topological surface state is observed due to its strong hybridization with the bulk bands. Near the $\overline{M}$ point, the two surface states, due to projections from two inequivalent bulk band inversions, interact and give rise to two peculiar sets of gapped surface states. The bulk band inversions and corresponding surface states can be tuned substantially by varying rare earth elements, in good agreement with density-functional theory calculations assuming local $f$ electrons. Our study therefore establishes rare earth mono-bismuthides as an interesting class of materials possessing tunable electronic properties and magnetism, providing a promising platform to search for novel properties in potentially correlated topological materials.

# I. INTRODUCTION

The discovery of topological insulator has sparked tremendous research interest in condensed matter physics in the past decade [1,2,3]. The recent realization of topological semimetal, including Dirac and Weyl semimetals [4,5,6,7,8,9,10,11,12], has further advanced the field and significantly expanded the scope of topological materials. While a majority of known topological materials are weakly correlated systems, searching for topological phases with strong electron correlations could yield a plethora of new quantum states and phenomena [13]. A well-known example of correlated topological material is the topological Kondo insulator, e.g., $SmB_6$, where the hybridization between $f$ and conduction electrons opens up an energy gap, which is smoothly connected by topological surface states (TSSs) [14,15,16,17,18]. Recently rare-earth (RE) mono-antimonides/bismuthides, RESb/REBi, have attracted considerable research interest as potential candidates for correlated topological semimetals [19,20,21,22]. In LaSb/LaBi where no $f$ electrons are present, it has been predicted that strong spin-orbit coupling (SOC) causes bulk band inversion, leading to nontrivial $Z_2$-topological invariant and TSSs [21]. Replacing La by other RE elements with partially filled $f$ shell, magnetic transition could occur at low temperature, which break time-reversal symmetry and potentially lead to novel topological phases with strong electronic correlations, such as Weyl Fermions [23]. The simultaneous presence of nontrivial band topology and magnetism provides an opportunity to study topological phase in a strongly correlated setting.

The electronic structure of RESb/REBi has been studied by first-principles calculations and angle-resolved photoemission spectroscopy (ARPES). Nontrivial band topology and TSSs have been observed experimentally in LaBi by ARPES measurements [24,25,26,27,28], while most RESb compounds are found to be topologically trivial without TSSs [29,30,31,32,33]. In this paper, we focus on the electronic structure of a few representative REBi compounds with different numbers of $f$ electrons, including CeBi, PrBi, SmBi and GdBi, in order to probe the evolution of bulk bands and possible TSSs. While most previous ARPES studies of REBi are focused on LaBi [24-28] (with two recent papers on CeBi [20,34]), such a systematic study is crucial to reveal the role of $f$ electrons and their influence to the electronic structure near the Fermi level.

## II. EXPERIMENTAL AND COMPUTATIONAL DETAILS

Single crystals of REBi were grown using an indium flux method with a molar ratio of RE: Bi: In of 1:1:10. The raw materials were weighed and put into $Al_2O_3$ crucible, sealed in an evacuated Quartz tube, and heated to 1100°C before cooling down slowly to 800°C. The excessive indium was removed in a centrifuge. The typical size of the crystal is 2x2x2 mm. After growth, the samples were characterized by Laue diffraction, magneto-transport and thermodynamic measurements. Synchrotron-based ARPES measurements were carried out at BL13U beamline at National Synchrotron Radiation Lab (NSRL, China) and beamline 21B at Taiwan Light Source, both equipped with R4000 electron energy analyzer. All ARPES spectra in this paper were taken at ~30 K, i.e., the samples are in their paramagnetic phase. The typical energy resolution is ~20 meV and the momentum resolution is typically 0.01 Å$^{-1}$. Constant energy contours are generated by rotating the sample in steps of 1° or 0.5° with respect to the vertical $z$ axis (in NSRL). Large single crystals of ReBi were cleaved *in-situ* in the ARPES chamber (base pressure ~7x10$^{-11}$ Torr), resulting in flat shinny surface for ARPES measurement. Because these samples are easily oxidized, all data presented in this paper were taken only within a few hours (<4 hours) after a fresh cleave at low temperature. Sample aging is monitored regularly (by checking the reference scan) to ensure that the data presented in the paper reflects the intrinsic band structure of the materials.

Electronic structure calculations were performed using density functional theory (DFT) and a plane-wave basis projected augmented wave method, as implemented in the Vienna Ab Initio Simulation Package (VASP) [35]. $f$ electrons were treated as core electrons in all DFT calculations. An energy cut-off of 480 eV and 12x12x12 Gamma-centered K-mesh were employed to converge the calculation to 1 meV/atom. Spin-orbit coupling effect were considered using a second variational step. Since the Perdew Burke and Ernzerhof (PBE) flavor of generalized gradient approximation to the exchange-correlation functional is known to exaggerate the band inversion features [36], we have employed the modified Becke-Johnson (MBJ) potential to obtain better agreement with experiments [37]. For the PBE calculations, the surface states were obtained by using a 25-layer slab-model with 25 Å vacuum layer (PBE slab); while for the MBJ calculations, the surface states were obtained with the surface Green's function method (MBJ GF calculations) [38], assuming a semi-infinite slab. Slab calculations using the MBJ potential are difficult to perform at the moment, due to the non-convergence problem.

## III. RESULTS AND DISCUSSIONS

### A. SAMPLE CHARACTERIZATION

REBi crystallizes in the simple rocksalt structure (face-centered cubic) as shown in Fig. 1(a), where its bulk 3D Brillouin zone (BZ) and projected 2D BZ, typically used for APRES analysis, are also shown. Since we work on the (001) surface in this paper, the bulk $\Gamma$ and one $X$ ($k_z=\pi$) point project onto 2D $\overline{\Gamma}$ point with ($k_x,k_y$)= (0,0), and two bulk symmetry-inequivalent $X$ points (with $k_z=0$ or $\pi$, respectively) project onto the same 2D $\overline{M}$ point with ($k_x,k_y$)= (~1 Å$^{-1}$,0).

Momentum-integrated energy scan reveals sharp core levels, including Bi $5d$ (–26 eV and -23 eV), RE $5p$ (between -28 and -18 eV) and RE $4f$ (from -10 eV up to -2 eV) (Fig. 1(b)), confirming good sample quality. The sample quality is further verified by the large residual resistivity ratio (typically over 100), as well as sharp Laue patterns, as shown in Fig. 1(c,d). An extremely large magnetoresistance (MR) was also observed [39] - an example is shown for CeBi with MR up to $2 \times 10^4$% at 9T and 2K. Magnetic transitions take place at low temperature for many REBi's, due to partially filled $4f$ shells, which may overlap with the orbits of conduction electrons and develop long-range magnetic order. A well-known example is CeBi, where highly anisotropic magnetic orderings with a complex phase diagram occur at low temperature, accompanied by dramatic changes in transport and thermodynamic properties [40,41,42,43]. While PrBi (non-Kramers $f^2$ configuration) stays in a paramagnetic phase at the lowest temperature, a magnetic transition takes place at low temperature for both SmBi ($f^5$) and GdBi ($f^7$). These magnetic transitions are evident in the resistivity data (Fig. 1(c)), where sharp change occurs near the anti-ferromagnetic (AFM) transition for CeBi ($T_N$=25 K) and SmBi ($T_N$=9K), with peak-like features typical for AFM Kondo lattice systems [44]. For GdBi ($T_N \simeq 27$ K), the AFM transition shows up as a small kink in the resistivity. In this paper, we focus on the basic electronic structure of REBi in the high temperature paramagnetic phase, where time-reversal symmetry is still preserved.

## B. BULK BAND INVERSION AND SURFACE STATES

In the paramagnetic phase, the calculated Fermi surface (FS) of bulk REBi consists of two hole pockets at the $\Gamma$ point (Bi $6p$ orbitals) and one electron pocket at each $X$ point (RE $5d$ orbitals), as illustrated in Fig. 1(e). Due to the three-dimensional nature of the band structure, we first performed a detailed photon energy dependence study to associate the APRES spectra with different $k_z$ cuts; the example of GdBi is shown in Fig. 2. The projected bulk band structure, obtained by DFT calculations using the MBJ potential, is also shown for comparison. It is clear that the spectra under 83 eV is very close to the $k_z = \pi$ cut (thick red curves in the calculations in Fig. 2(g)), showing two cone-like structures at $\overline{\varGamma}$ and $\overline{M}$ point, respectively. The 54 eV spectra shows strong emission from hole bands near $\overline{\varGamma}$ point and electron bands close to the $\overline{M}$ point, corresponding to the $k_z = 0$ cut (thick black curves in Fig. 2(g)). Based on the photon energy vs $k_z$ correspondence at these photon energies, one could estimate the inner potential to be $\simeq 14$ eV, similar to previous results for LaBi [25]. We should mention that significant $k_z$ broadening is present in REBi; as a result, the bulk bands do not move continuously with photon energy as one might expect for a bulk band under the standard dipole transition. For example, the bulk hole pockets at $\overline{\varGamma}$ point are observed over a large photon energy range between the "expected" $k_z \sim 3\pi$ (23.8 eV) and $k_z \sim 4.5\pi$ (65 eV), and the bands show little movement with photon energy (cf. the dashed yellow curves). Such a discontinuous movement of bulk bands with photon energy can be attributed to a large $k_z$ broadening, which has also been observed in other mono-pnictide systems, such as LaSb, CeSb [26,32,34]. The most likely cause for the large $k_z$ broadening is the short photoelectron escape depth (5-10 Å) in the current experiment; Indeed, soft X-ray ARPES in CeSb/CeBi has identified continuous movement of bulk bands with photon energies, due to better bulk sensitivity (hence smaller $k_z$ broadening) [34]. The large $k_z$ broadening in the current photon energy range indicate that it is necessary to make careful comparison between experiments and calculations to distinguish bulk bands and TSSs.

According to DFT calculations, bulk band inversion (between Bi $6p$ and RE $5d$ bands) takes place along $\varGamma X$ direction, as indicated by Fig. 2(h), and leads to nontrivial $Z_2$-topological invariant and TSSs [24]. Here the $Z_2$-topological invariant can be evaluated by calculating the band parity of all occupied bands at all time-reversal invariant momentum points [45]. Although the definition of $Z_2$-topological invariant was intended for topological insulator with an absolute band gap, such concept can be extended to semi-metallic REBi, where a momentum-dependent partial band gap exists, separating the valence and conduction bands [12]. Detailed numerical calculations predict that REBi (with RE ranging from Ce to Gd) could possess nontrivial $Z_2$ index and hence TSSs [46]. Since there are three $X$ points in this calculation, the $Z_2$ index is dependent on the product of the parity of all occupied bands at $X$: if there is band inversion along the $\Gamma X$ direction, the $Z_2$-topological invariant is nontrivial; otherwise, the system will be topologically trivial. This bulk band inversion is indeed observed in all compounds studied in this paper; an example for PrBi is shown in Fig. 3 (same for GdBi in Fig. 2). The bulk band inversion is best displayed under 54 eV photons (Fig. 3(a)), where the Pr 5d (black dashed curve) and Bi 6p (red dashed curve) bands become crossed and yield an inverted gap, in good agreement with MBJ GF calculation (the rightmost panel). The emission intensity of bulk bands do not show obvious shift across the inversion point, which seems strange in view of the changed orbital character. However, it is likely that the photoexcited final state plays an important role in the photoemission matrix element, which, together with the large $k_z$ broadening, smears out the expected intensity shift across the inversion point.

In the MBJ GF calculation, two sets of gapless TSS Dirac cones would be expected at the $\overline{M}$ point due to projection of two symmetry-inequivalent $X$ points, which consists of linearly dispersive Dirac cones and relatively flat portions that smoothly connect to the bulk inversion point. In experiments (23.8 and 28.1 eV data), we could only resolve the lower portion of Dirac cone, while the upper part is somewhat diffuse and cannot be identified clearly. As we shall see later (Section c), two predicted TSSs near the $\overline{M}$ point (based on projections of two bulk $X$ points) actually interact with each other and give rise to two sets of gapped surface states (SSs). As these SSs disperse into bulk band region, it gradually loses its surface character and merges into the bulk continuum. Although the SSs presumably share similar in-plane orbital character as the bulk states, their detailed photon energy dependence seems to be obviously different, likely due to the difference in the $z$ extent of the wavefunction (hence a different photoexcitaton cross section).

The experimental FS (Fig. 3(b)) at $\overline{\Gamma}$ point consists of two hole pockets ($k_z$=0) and one electron pocket ($k_z$=π), due to a large $k_z$ broadening as mentioned earlier. Two elliptically shaped electron pockets, rotated 90° with respect to one another, can be observed at the $\overline{M}$ point, corresponding to $k_z$=0 and π cuts respectively. All bands at the FS are expected to derive from bulk states. Moving down in energy, the hole pockets expand and the electron pockets shrink; at the bulk band inversion point ($E \simeq$ -0.2 eV), one begins to see significant spectral contributions from SSs near the $\overline{M}$ point, which exhibit elliptical shape and eventually develops continuous connection to bulk bands with cross-like features ($E \simeq$ -0.25 eV). The highly anisotropic dispersion of the SSs is a direct manifestation of the bulk band structure. All these experimental observations are in good agreement with MBJ GF calculations (Fig. 3(c)). Note that the MBJ GF calculations assume equal spectral contributions from all $k_z$'s; the good agreement between ARPES results and MBJ GF calculations further confirms the large $k_z$ broadening inherent in the photoemission process of REBi.

A $Z_2$ topological insulator is characterized by odd number of gapless TSS(s), and, in the current case, it should theoretically correspond to a gapless TSS at the $\overline{\Gamma}$ point (due to the projection of a single $X$ point). This predicted TSS stems from the same bulk band inversion observed near the $\overline{M}$ point, except that it is now from the $k_z$~π projection (see Fig. 1(a)). The ARPES spectra near the $\overline{\Gamma}$ point for CeBi, PrBi, SmBi and GdBi are summarized in Fig. 4(a).

Indeed, a cone-like feature can be observed near the $\overline{\Gamma}$ point for all compounds, which extends in a large energy range and forms a continuous dispersion across the valence band. In addition, this sharp cone-like feature is present over a large photon energy range (see Fig. 5). Although these behaviors seem to be compatible with the theoretically expected gapless TSS, its dispersion with photon energy in some cases (e.g., in Fig. 2(a,b)) suggests that it is more likely due to the neighboring bulk states near $k_z \sim \pi$. We argue that due to the absence of a partial bulk gap near the $\overline{\Gamma}$ point, the originally gapless TSS hybridizes strongly with the neighboring bulk bands and therefore losses its surface character. This might be why this Dirac-cone-like feature could appear gapped in some cases [25,27], due to complete overlap with gapped bulk states. This also highlights the unique character of a semimetal with $Z_2$-topological invariant: even if the bulk topology predicts odd number of gapless TSS(s), some TSS(s) could hybridize strongly with neighboring bulk states and no longer survive as pure SSs.

## C. EVOLUTION OF ELECTRONIC STRUCTURE WITH RARE EARTH ELEMENTS

With heavier RE's, the cone-like features near the $\overline{\Gamma}$ point move down in energy (see white arrows in Fig. 4(a)), while the bulk hole bands show relatively little change, consistent with the MBJ GF calculations (Fig. 4(b)). In the calculations, a bundle of bulk states piles up near the predicted TSS, which could give rise to visually "gapped Dirac cones" centered at the $\overline{\Gamma}$ point.

In Figure 6(a,b,c), we present the evolution of the bulk bands and associated SSs near the $\overline{M}$ point for CeBi, SmBi and GdBi, respectively. Moving from Ce to Gd, we observe that the inverted bulk gap (denoted by cyan arrows in Fig. 6(c)) gradually shrinks and the band inversion region moves down in the energy, in good agreement with the MBJ GF calculations (Fig. 6(d)). Interestingly, both experiments and calculations indicate that GdBi possesses a very small inverted gap and highly linear bulk bands, suggesting a possible topologically nontrivial-trivial phase transition when replacing Gd with a slightly heavier RE element. Specifically, DFT calculations have predicted that no bulk band inversion occurs for REBi with RE heavier than Dy [46], resulting in a topologically trivial phase. The large tunability of both bulk bands and SSs in REBi is quite different to the case of RESb, where the change of the electronic structure with RE replacement is much less pronounced [31].

While the experimental bulk bands are in good agreement with MBJ GF calculations, the detailed dispersion of SSs near the $\overline{M}$ point differs. The experimentally extracted dispersions of SSs are highlighted by yellow dashed lines in Fig. 6(a), which clearly show two cones with well separated Dirac points, apparently different from the MBJ GF calculations (Fig. 6(d)). The difference is probably due to the simplified treatment of surfaces in the GF calculations, which neglects the detailed surface conditions, particularly surface structural distortions and dangling bonds. This discrepancy could be solved by slab calculations that take into account of realistic surface conditions, but such slab calculations using the MBJ potential are hammered by the non-convergence problem in the practical calculation. To circumvent this problem, we present the results of slab calculations using the PBE potential in Fig. 6(e). Indeed, the results show two sets of vertically separated SSs, both of which consist of flat portions connecting to the band inversion point and V shaped cones centered at the $\overline{M}$ point, in better agreement with experimental data. Nevertheless, PBE slab calculations yield band inversions that are much larger than experiments, a well-known problem for DFT calculations using the PBE potential [47]. Note that the observed SSs are not real TSSs, since they are apparently gapped and connect to valence band or conduction band separately, different from intrinsic TSSs that form a continuous connection between valence and conduction bands. Nevertheless, these SSs are thought to be intimately linked to nontrivial band inversion near the *X* point, as no SS could be seen in the calculations of heavy REBi where no band inversion occurs [46]. A possible explanation is that a direct projection of two symmetry-inequivalent *X* points give rise to two sets of originally gapless TSSs, which then interact and open up a gap, rendering two sets of trivial gapped SSs. The gap opening might be related to surface structural distortions, which act to reduce the dangling bonds at the surface and minimize the surface energy. These SSs could also be interpreted in terms of classic Shockley-type SSs, which arise due to closings and openings of local band gaps [48].

To understand the dependence of the electronic structure on the RE element, one needs to disentangle the effects of the varied proton/*f* - electron counting and simultaneous lattice compression. This insight would be useful to understand the pressure tuning of physical properties in REBi [49]. To simulate pure lattice-pressure effect, we have calculated the electronic structure of CeBi with differing lattice constants, matched to bulk CeBi, SmBi and GdBi, respectively (Fig. 7). Apparently, pure lattice compression enlarges the bulk gap inversion due to the stronger wave function overlap, which is opposite to the experimental trend of replacing with heavier RE elements (Fig. 6(a)). This indicates that the varied proton/*f* - electron counting plays a dominant role in tuning the electronic structure, which essentially changes the atomic and exchange potential that the valence/conduction electrons feel. Note that *f* electrons should be mostly local in the paramagnetic phase, yet they could not be simply considered as partially cancelling the nuclei charge.

## IV. CONCLUDING REMARKS

To conclude, we present ARPES measurements and comparison with electronic structure calculations for REBi homologues, a class of materials possessing possibly nontrivial bulk band topology and strong electronic correlations below magnetic transitions. We identified the bulk band inversion, in excellent agreement with MBJ calculations. We did not observe clear gapless TSS near the $\bar{\Gamma}$ point, probably due to strong mixing with neighboring bulk bands, while two of the originally gapless TSSs near the $\bar{M}$ point interact and yield a peculiar pair of gapped trivial SSs. The large variation of the electronic structure with RE elements provides an opportunity to tune the topological properties, while the presence of magnetism at low temperature sets the stage for exploring strongly correlated topological phases.

Our current results call for future studies to explore the effect of magnetism on the electronic structure at low temperature. Specifically, it would be important to track the change of electronic structure across the magnetic transition and understand the role of the $f$ electrons. It might also be interesting to reveal SSs under different magnetically ordered phases and search for novel topological phases with broken time-reversal symmetry. Experimental efforts are currently under way to tackle these problems.

This work is supported by National Key R&D Program of the MOST of China (Grant No. 2017YFA0303100, 2016YFA0300203, 2014CB648400), National Science Foundation of China (No. 11674280, No. 11274006), the Science Challenge Program of China. TCC acknowledges support from the US Department of Energy under Grant No. DE-FG02-07ER46383. We would like to thank Mr. Pengdong Wang, Dr. Chanyuen Chang for support and help during synchrotron ARPES measurements, and Prof. Fuchun Zhang, Prof. Haijun Zhang, Prof. Stefan Kirschner for helpful discussions.

## References

[1] M. Z. Hasan, and C. L. Kane, Rev. Mod. Phys. **82**, 3045 (2010).

[2] X.-L. Qi, and S.-C. Zhang, Rev. Mod. Phys. **83**, 1057 (2011).

[3] M. König, S. Wiedmann, C. Brüne, A. Roth, H. Buhmann, L.-W.Molenkamp, X.-L.Qi and S.-C. Zhang, Science **318**, 766 (2007).

[4] Z. K. Liu, B. Zhou, Y. Zhang, Z.-J. Wang, H.-M. Weng, D. Prabhakaran, S.-K. Mo, Z.-X. Shen, Z. Fang, X. Dai, Z. Hussain and Y.-L. Chen, Science **343**, 864 (2014).

[5] S.-Y. Xu, C. Liu, S. K. Kushwaha, R. Sankar, J. W. Krizan, I. Belopolski, M. Neupane, G. Bian, N. Alidoust, T.-R. Chang. H.-R. Jeng, C.-Y. Huang, W.-F. Tsai, H. Lin, P.-P. Shibayev, F.-C. Chou, R.-J. Cava and M. Zahid Hasan, Science **347**, 294 (2015).

[6] X. Wan, A. M. Turner, A. Vishwanath and S. Y. Savrasov, Phys. Rev. B **83**, 205101 (2011).

[7] S.-M. Huang, S.-Y. Xu, I. Belopolski, C.-C. Lee, G.-Q. Chang, B.-K. Wang, N. Alidoust, G. Bian, M. Neupane, C.-L. Zhang, S. Jia, A. Bansil and M. Zahid Hassan, Nat. Commun. **6**, 7373 (2015).

[8] H. Weng, C. Fang, Z. Fang, B. A. Bernevig and X. Dai, Phys. Rev. X **5**, 011029 (2015).

[9] L. X. Yang, Z. K. Liu, Y. Sun, H. Peng, H. F. Yang, T. Zhang, Y. F. Guo, M. Rahn, D. Prabhakaran, Z. Hussain, S.-K. Mo, C. Felser, B. Yan and Y. L. Chen, Nat. Phys. **11**, 728 (2015).

[10] B. Q. Lv, H. M. Weng, B. B. Fu, X. P. Wang, H. Miao, J. Ma, P. Richard, X. C. Huang, L. X. Zhao, G. F. Chen, Z. Fang, X. Dai, T. Qian, and H. Ding, Phys. Rev. X **5**, 031013 (2015).

[11] S.-Y. Xu, I. Belopolski, N. Alidoust, M. Neupane, G. Bian, C.-L. Zhang, R. Sankar, G.-Q. Chang, Z.-J. Yuan, C.-C. Lee, S.-M. Huang, H. Zheng, J. Ma, Daniel S. Sanchez, B.-K. Wang, A. Bansil, F.-C. Chou, Pavel P. Shibayev, H. Lin, S. Jia and M. Zahid Hassan, Science **349**, 613 (2015).

[12] H. Weng, X. Dai and Z. Fang, J. Phys.: Condens. Matter **28**, 303001 (2016).

[13] X. Zhang, H.-J. Zhang, C. Felser, S.-C. Zhang, Science **335**, 1464 (2012).

[14] M. Dzero, K. Sun, V. Galitski and P. Coleman, Phys. Rev. Lett. **104**, 106408 (2010).

[15] X.-H. Zhang, N. P. Butch, P. Syers, S. Ziemak, Richard L. Greene, and J. Paglione, Phys. Rev. X **3**, 011011 (2013).

[16] N. Xu, X. Shi, P. K. Biswas, C. E. Matt, R. S. Dhaka, Y. Huang, N. C. Plumb, M. Radović, J. H. Dil, E. Pomjakushina, K. Conder, A. Amato, Z. Salman, D. M. Paul, J. Mesot, H. Ding, and M. Shi, Phys. Rev. B **88**, 121102(R) (2013).

[17] M. Neupane, N. Alidoust, S.-Y. Xu, T. Kondo, Y. Ishida, D. J. Kim, C. Liu, I. Belopolski, Y. J. Jo, T.-R. Chang, H.-T. Jeng, T. Durakiewicz, L. Balicas, H. Lin, A. Bansil, S. Shin, Z. Fisk and M. Z. Hassan, Nat. Commun. **4**:2991 (2013).

[18] J. Jiang, S. Li, T. Zhang, Z. Sun, F. Chen, Z. R. Ye, M. Xu, Q. Q. Ge, S. Y. Tan, X. H. Niu, M. Xia, B. P. Xie, Y. F. Li, X. H. Chen, H. H. Wen and D. L. Feng, Nat. Commun. **4**:3010 (2013)

[19] F. F. Tafti, Q. D. Gibson, S. K. Kushwaha, N. Haldolaarachchige, and R. J. Cava, Nat. Phys. **12**, 272 (2015).

[20] N. Alidoust, A. Alexandradinata, S.-Y. Xu, I. Belopolski, S. K. Kushwaha, M.-G. Zeng, M. Neupane, G. Bian, C. Liu, D. S. Sanchez, P. P. Shibayev, H. Zheng, L. Fu, A. Bansil, H. Lin, R. J. Cava and M. Z. Hassan, arXiv:1604.08571.

[21] M. Zeng, C. Fang, G.-Q. Chang, Y.-A. Chen, T. Hsieh, A. Bansil, H. Lin, and L. Fu, arXiv:1504.03492.

[22] S. Jang, R. Kealhofer, C. John, S. Doyle, J. Hong, J.-H. Shim, Q. M. Si, O. Erten, J. D. Denlinger, and J. G. Analytis, arXiv:1712.05817v1.

[23] C. Y. Guo, C. Cao, M. Smidman, F. Wu, Y.-J. Zhang, F. Steglich, F.-C. Zhang, and H. Q. Yuan, NPJ Quantum Materials **2**, 39 (2017).

[24] J. Nayak, S.-C. Wu, N. Kumar, C. Shekhar, S. Singh, J. Fink, E. Rienks, G. H. Fecher, S. Parkin, B. Yan, and C. Felser, Nat. Commun. **8**:13942 (2016).

[25] R. Lou, B.-B. Fu, Q. N. Xu, P.-J. Guo, L.-Y. Kong, L.-K. Zeng, J.-Z. Ma, P. Richard, C. Fang, Y.-B. Huang, S.-S. Sun, Q. Wang, L. Wang, Y.-G. Shi, H.C. Lei, K. Liu, H. M. Weng, T. Qian, H. Ding, and S.-C. Wang, Phys. Rev. B **95**, 115140 (2017).

[26] X. H. Niu, D. F. Xu, Y. H. Bai, Q. Song, X. P. Shen, B. P. Xie, Z. Sun, Y. B. Huang, D. C. Peets, and D. L. Feng, Phys. Rev. B **94**, 165163 (2016).

[27] Y. Wu, T. Kong, L.-L. Wang, D. D. Johnson, D.-X. Mou, L. Huang, B. Schrunk, S. L. Bud'ko, P. C. Canfield, and A. Kaminski, Phys. Rev. B **94**, 081108 (R) (2016).

[28] B. Feng, J. Cao, M. Yang, Y. Feng, S. Wu, B. Fu, M. Arta, K. Miyamoto, S. He, K. Shimada, Y. Shi, T. Okuda, and Y.-G. Yao, Phys. Rev. B **97**, 155153(2018).

[29] L.-K. Zeng, R. Lou, D.-S. Wu, Q. N. Xu, P.-J. Guo, L.-Y. Kong, Y.-G. Zhong, J.-Z. Ma, B.-B. Fu, P. Richard, P. Wang, G. T. Liu, L. Lu, Y.-B. Huang, C. Fang, S.-S. Sun, Q. Wang, L. Wang, Y.-G. Shi, H. M. Weng, H.-C. Lei, K. Liu, S.-C. Wang, T. Qian, J.-L. Luo, and H. Ding, Phys. Rev. Lett. **117**, 127204 (2016).

[30] J. He, C.-F. Zhang, N. J. Ghimire, T. Liang, C.-J. Jia, J. Jiang, S.-J, Tang, S. Chen, Y. He, S.-K. Mo, C/ C. Hwang, M. Hashimoto, D. H. Lu, B. Moritz, T. P. Deveraux, Y. L. Chen, J. F. Mitchell, and Z.-X. Shen, Phys. Rev. Lett. **117**, 267201 (2016).

[31] Y. Wu, Y. Lee, T. Kong, D. X. Mou, R. Jiang, L. Huang, S. L. Bud'ko, P. C, Canfield, and A. Kaminski, Phys. Rev. B **96**, 035134 (2017).

[32] H. Oinuma, S. Souma, D. Takane, T. Nakamura, K. Nakayama, T. Mitsuhashi, K. Horiba, H. Kumigashira, M. Yoshida, A. Ochiai, T. Takahashi, and T. Sato, Phys. Rev. B **96**, 041120 (R) (2017).

[33] M. Neupane, M. M. Hosen, I. Belopolski, N. Wakeham, K. Dimitri, N. Dhakal, J.-X. Zhu, M. Z. Hasan, E. D. Bauer, and F. Ronning, J. Phys. Condens. Matter **28**, 23LT02 (2016).

[34] K. Kuroda, M. Ochi, H. S. Suzuki, M. Hirayama, R. Noguchi, C. Bareille, S. Akebi, S. Kunisada, T. Muro, M. D. Watson, H, Kitazawa, Y. Haga, T. K. Kim, M. Hoesch, S. Shin, R. Arita, and T. Kondo, Phys. Rev. Lett. **120**, 086402 (2018).

[35] G. Kresse and J. Hafner, Phys. Rev. B **47**, 558 (1993)

[36] J. P. Perdew, K. Burke and M. Ernzerhof, Phys. Rev. Lett. **77**, 3865 (1996)

[37] F. Tran and P. Blaha, Phys. Rev. Lett. **102**, 226401(2009)

[38] M. P. L. Sancho, J. M. L. Sancho, J. M. L. Sancho and J. Rubio, J. Phys. F **15**, 851 (1985)

[39] F.-F. Tafti, Q. Gibson, S. Kushwaha, J. W. Krizan, N. Haldolaarachchige, and R. J. Cava, Proc. Nat. Aca. Sci. **113**, E3475 (2016)

[40] F. Hulliger, M. Landolt, H. R. Ott, and R. Schmelczer, J. Low Temp. Phys. **20**, 269 (1975).

[41] H. Bartholin, D. Florence, Tcheng-Si Wang, and O. Vogt, Phys. Stat. Sol. **24**, 631 (1974).

[42] H. Bartholin, P. Burlet, S. Quezel, J. Rossat-Mignod and O. Vogt, J. Physique 40, C5 130 (1979).

[43] H. Takahashi and T. Kasuya, J. Phys. C: Solid State Phys. **18**, 2697 (1985).

[44] P. Coleman, Heavy Fermions: electrons at the edge of magnetism. Handbook of Magnetism and Advanced Magnetic Materials, vol. 1 (Wiley, New York, 2007).

[45] L. Fu and C. L. Kane, Phys. Rev. B **76**, 045302 (2007).

[46] X. Duan, F. Wu, J. Chen, P. Zhang, Y. Liu, H. Yuan, and C. Cao, arXiv: 1802.04554.

[47] P. J. Guo, H. C. Yang, B. J. Zhang K. Liu, and Z. Y. Lu, Phys. Rev. B **93**, 235142 (2016).

[48] W. Shockley, Phys. Rev. **56**, 317 (1939).

[49] F. F. Tafti, M. S. Torikachvili, R. L. Stillwell, B. Baer, E. Stavrou, S. T. Weir, Y. K. Vohra, H.-Y. Yang, E. F. McDonnell, S. K. Kushwaha, Q. D. Gibson, R. J. Cava, and J. R. Jeffries, Phys. Rev. B **95**, 014507 (2017).

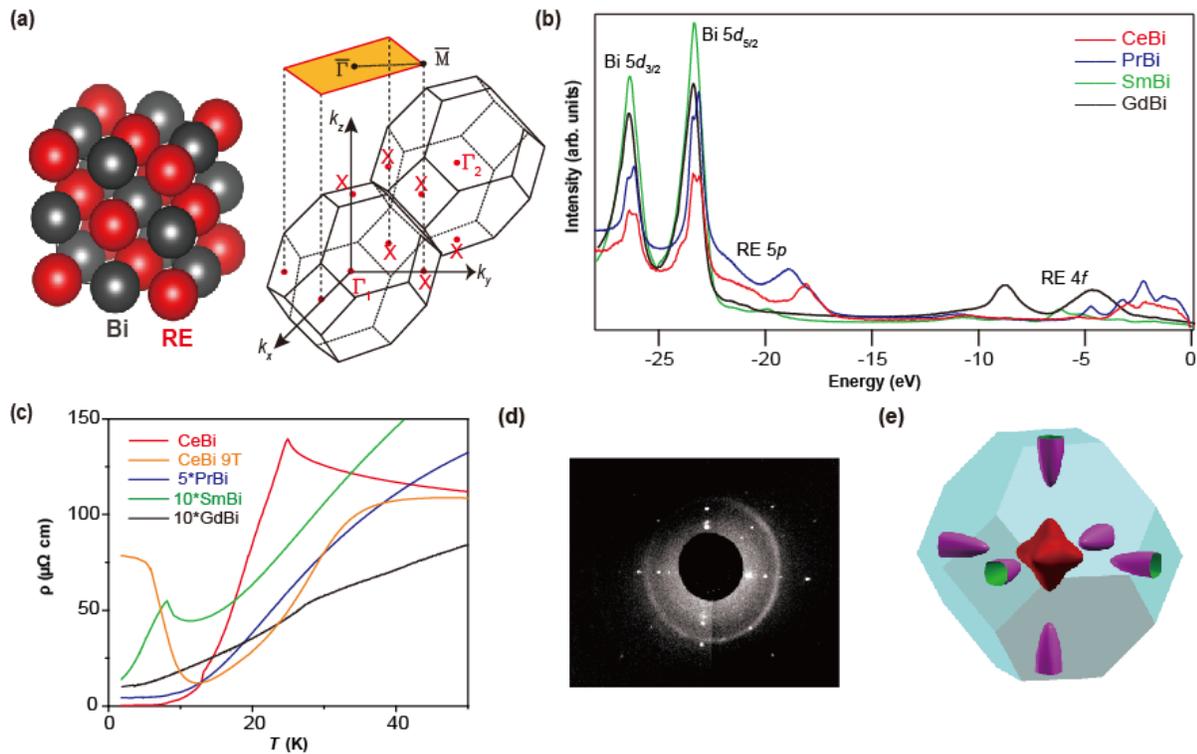

Fig. 1 (Color online). Sample characterization of REBi. (a) Crystal structure (left) and its corresponding bulk and surface BZ (right) of REBi. (b) Momentum-integrated large-range energy scans for various REBi compounds. The spectrum for CeBi, PrBi was taken with 42 eV photons, while 54 eV photons were used for SmBi and GdBi. The small splitting of both the Bi 5d$_{3/2}$ and 5d$_{5/2}$ lines is caused by simultaneous contributions from surface and bulk Bi atoms due to the cleaved surface, and their relative ratio could be dependent on the photon energy. The spectra between -2 and -10 eV are dominated by RE 4f bands, which gradually move to deeper binding energies with heavier RE. (c) The resistivity vs temperature for all REBi compounds. The resistivity for CeBi under 9T magnetic field is also displayed, showing its extremely large magnetoresistance. (d) A Laue image of SmBi indicating high quality single crystal. (e) Three-dimensional FS for GdBi within the first BZ, typical for all REBi's.

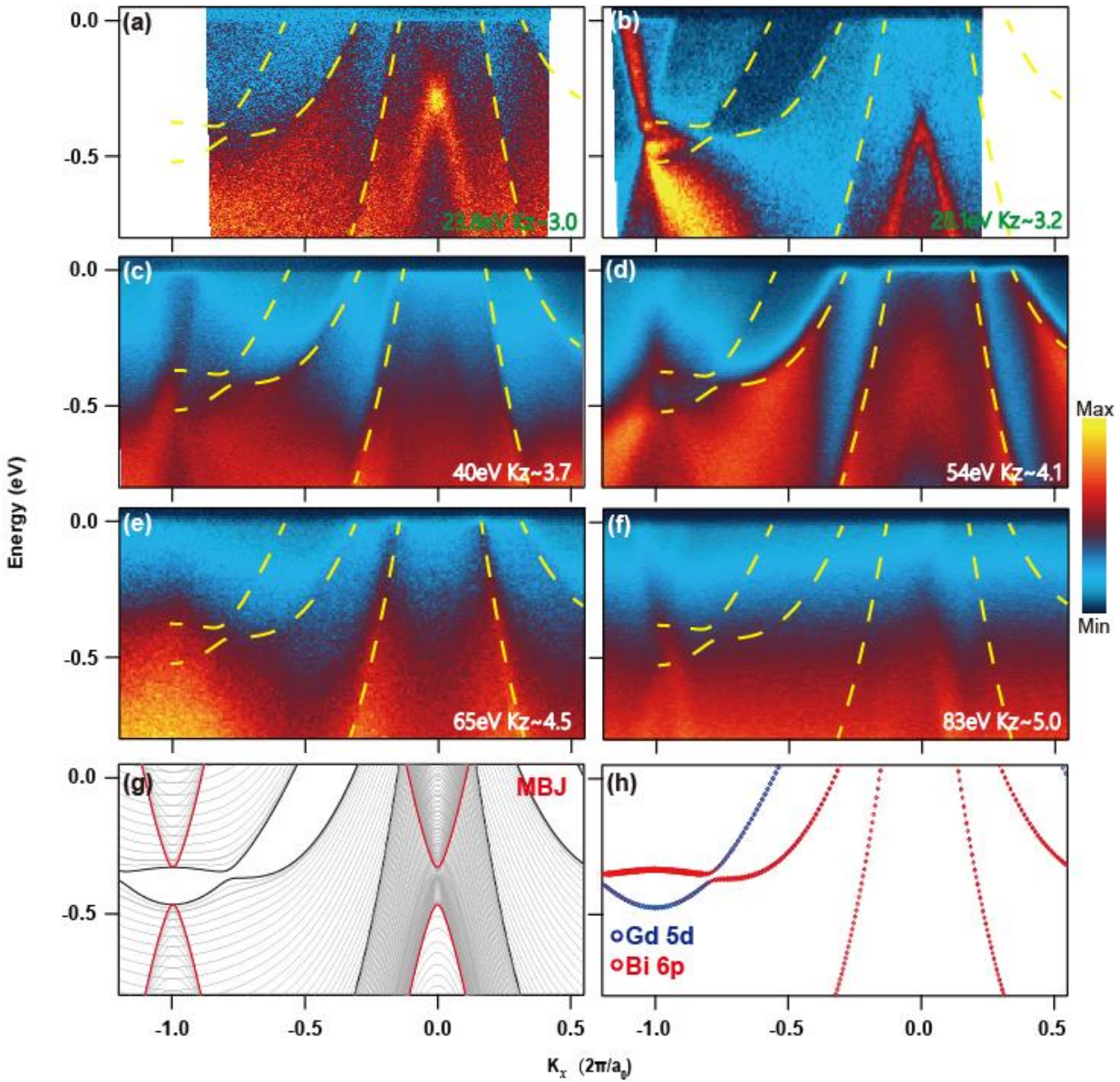

Fig. 2 (Color online). Probing the three-dimensional band structure by photon-energy dependent ARPES, example of GdBi. (a-f) ARPES spectra of GdBi taken along $\overline{\Gamma M}$ direction with various photon energies, in comparison with the projected bulk band calculation using the MBJ potential (g). The $k_z$ values, indicated at the bottom right of the experimental data, are calculated based on the dipole transition model using an estimated inner potential of 14 eV. The dashed yellow curves on top of experimental data are the extracted dispersion relations of the bulk bands at $k_z$=0 cut. The fact that they do not show continuous change with photon energy indicates large $k_z$ broadening in the photoemission process, which yields emission features dominated by the band edge ($k_z$=0 or π cuts). The thick black (red) curves in (g) indicates $k_z$=0 ($k_z$=π) cuts. (h) The band character for the $k_z$=0 cut, highlighting the bulk band inversion between Gd 5d and Bi 6p bands near $\overline{M}$ point.

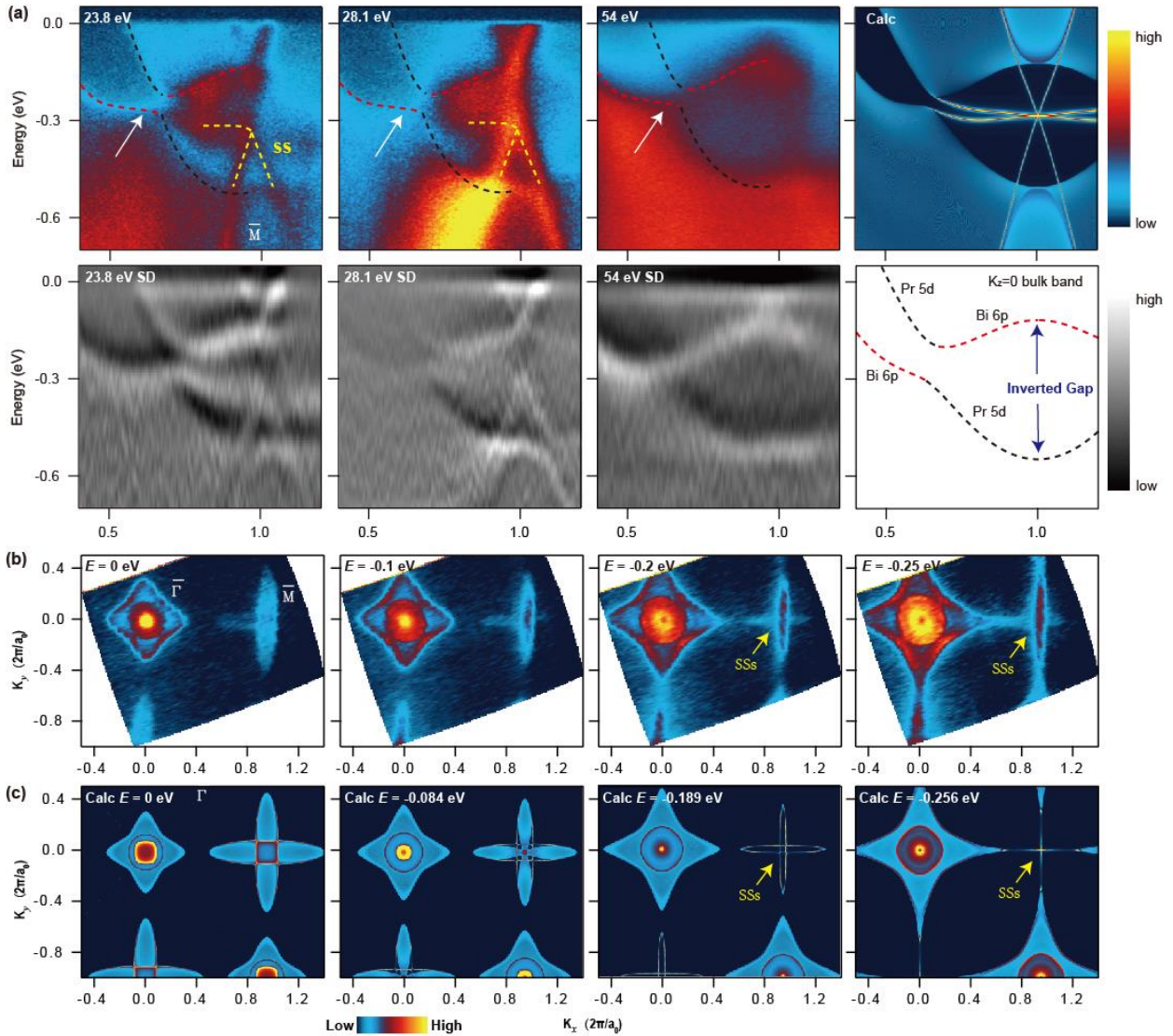

Fig. 3 (Color online). Identifying bulk band inversion and associated SSs, using PrBi data as an example. (a) Upper row: APRES data along $\overline{\Gamma M}$ direction taken with different photon energies, compared with the calculation (the rightmost panel). The yellow, black and red dashed curves indicate extracted SSs, bulk band edge of Pr 5d and Bi 6p, respectively. The white arrows indicate the bulk band inversion point. In the calculation, the blue background indicates the bulk continuum and the SSs are shown as sharp lines lying within the inverted gap. Lower row: the corresponding second derivative of experimental data, together with the plot of bulk band character for $k_z$=0 cut (the rightmost panel), which is directly relevant for bulk band inversion. (b) Experimental constant energy contours using 23.8 eV photons at $E$=0, -0.18, -0.24 and -0.315 eV, which corresponds to the Fermi energy, the conduction band bottom, the bulk band inversion point and the center point of SSs. (c) Theoretical constant energy contours, to be compared with (b). Thin lines at $E$=-0.189 and -0.256 eV maps corresponds to SS contours centered at the $\overline{M}$ point. All calculations were performed using the MBJ GF method.

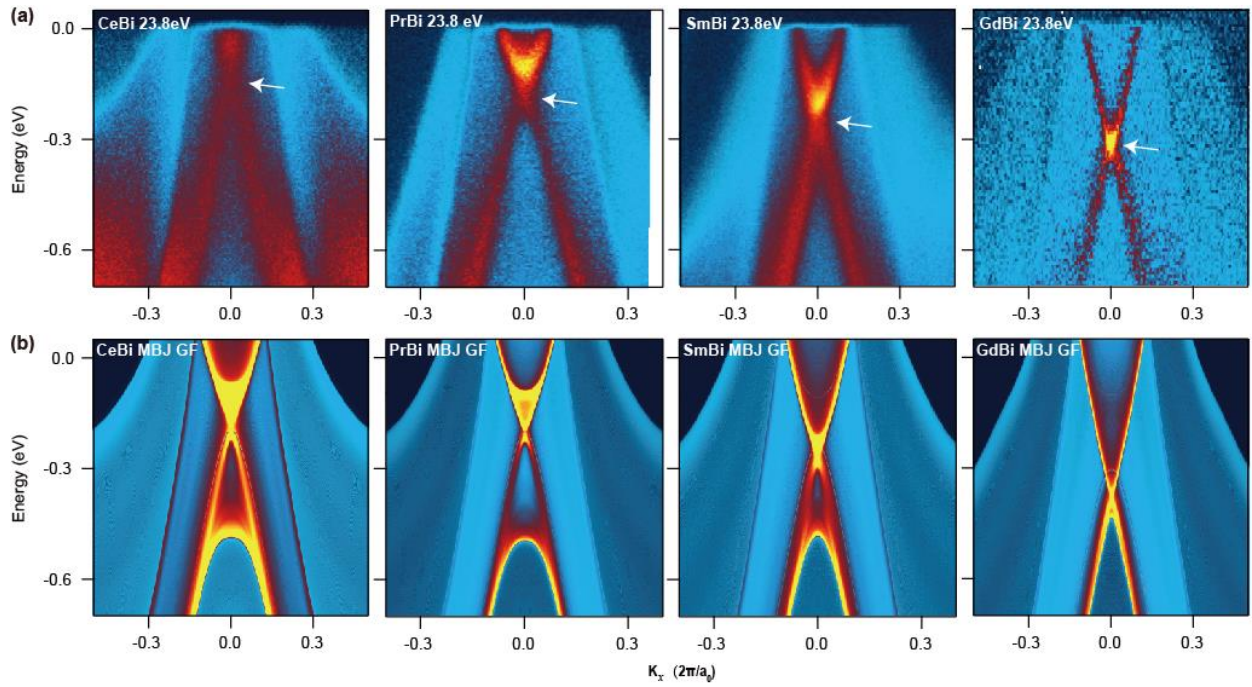

Fig. 4 (Color online). ARPES spectra near the $\overline{\varGamma}$ point for CeBi, PrBi, SmBi and GdBi, in comparison with theoretical calculations. (a) Experimental band dispersion near the $\overline{\varGamma}$ point taken at 23.8 eV. The white arrows indicate the crossing points of the cone-like emission features. (b) Results from MBJ GF calculations, to compare with experiments.

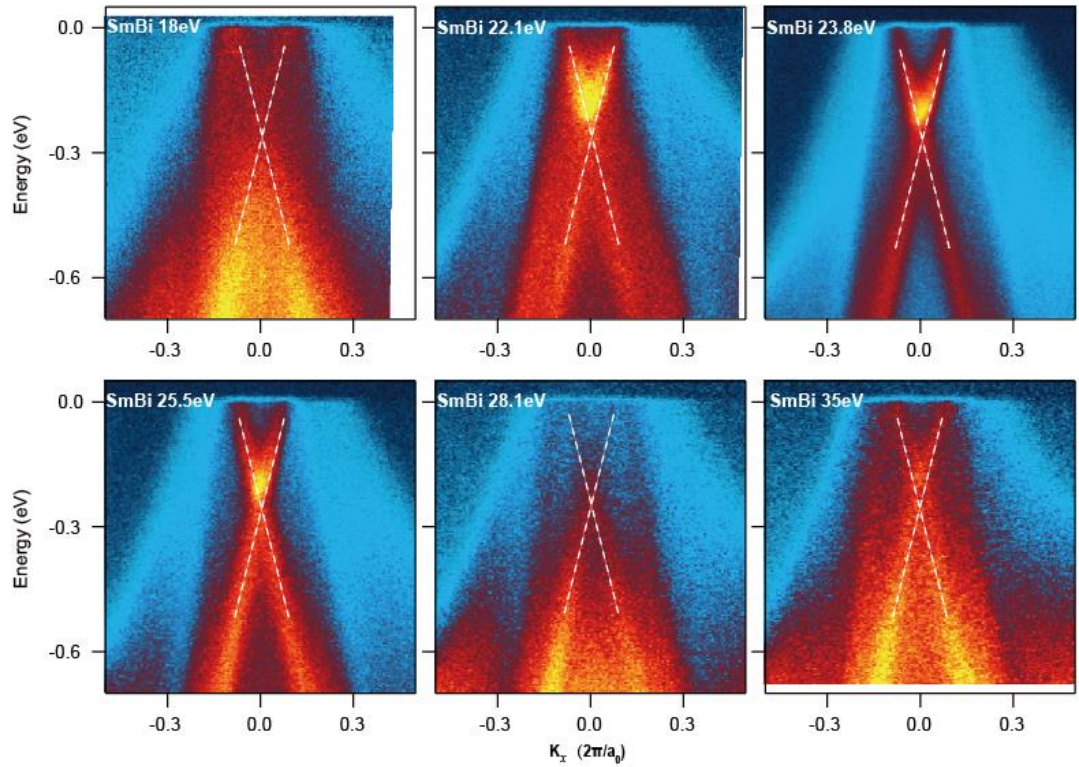

Fig. 5 (Color online). Photon energy dependent ARPES spectra of SmBi near the $\overline{\Gamma}$ point, showing a cone-like emission feature (denoted by the white dashed curves), which is visible over a wide photon energy range and does now show obvious change with photon energy. One possible explanation for this cone-like feature is the predicted gapless TSS, although it could be more naturally explained by the bulk bands near band inversion point, due to the large $k_z$ broadening discussed in the manuscript.

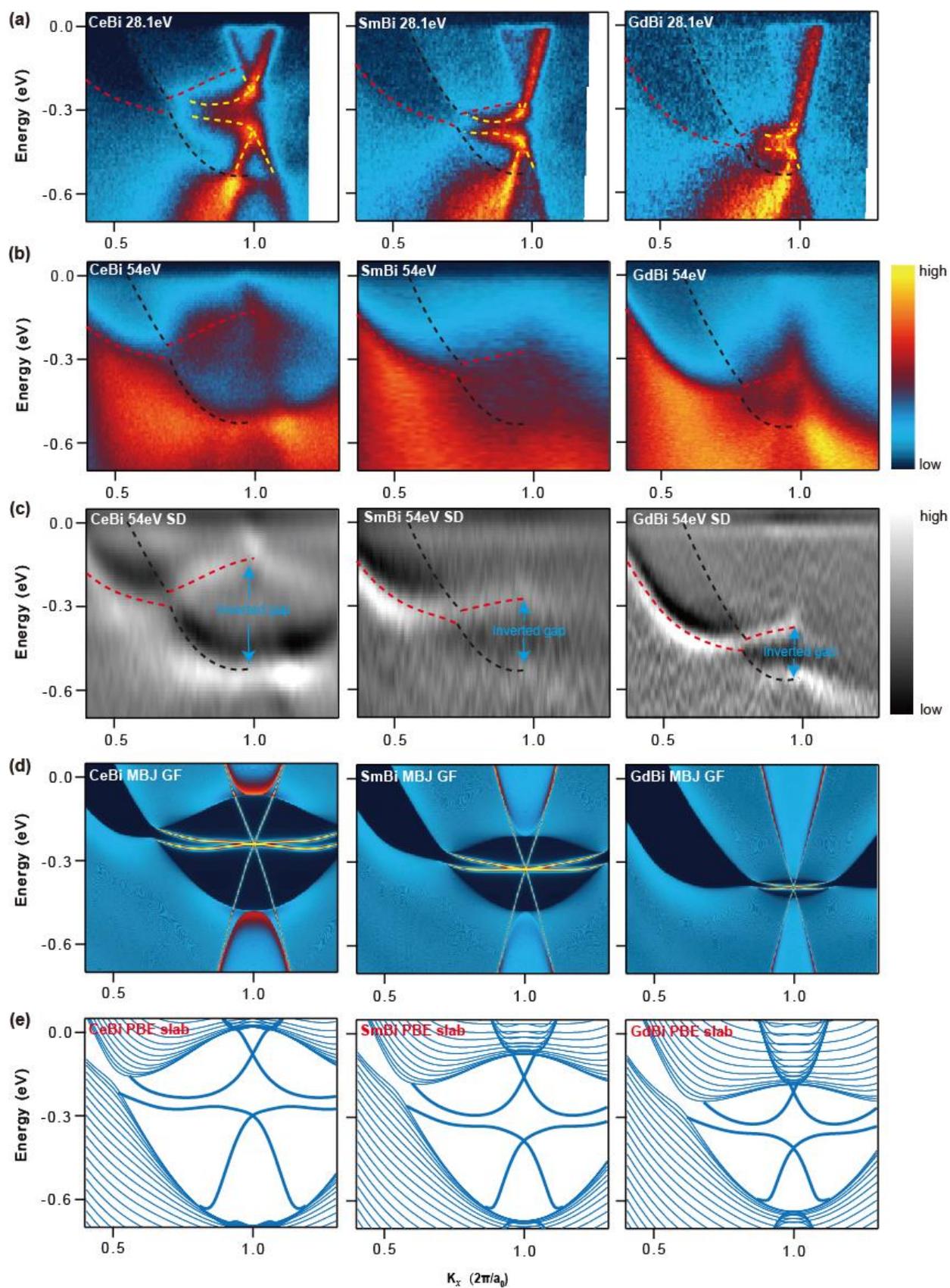

Fig. 6 (Color online). Evolution of bulk band inversion and associated SSs in CeBi, SmBi and GdBi, in comparison with theoretical calculations. (a,b) Experimental band dispersion along $\overline{\Gamma}\overline{M}$ direction taken at two representative photon energies. (c) is the second derivative of the data taken at 54 eV. The yellow dashed lines are the extracted dispersion of the SSs, while the black (red) dashed lines indicate bulk band edge of RE 5d (Bi 6p) band. The inverted bulk gap at the $\overline{M}$ point is indicated by cyan arrows in (c). (d,e) Results from MBJ GF calculations (d) and PBE slab calculations (e).

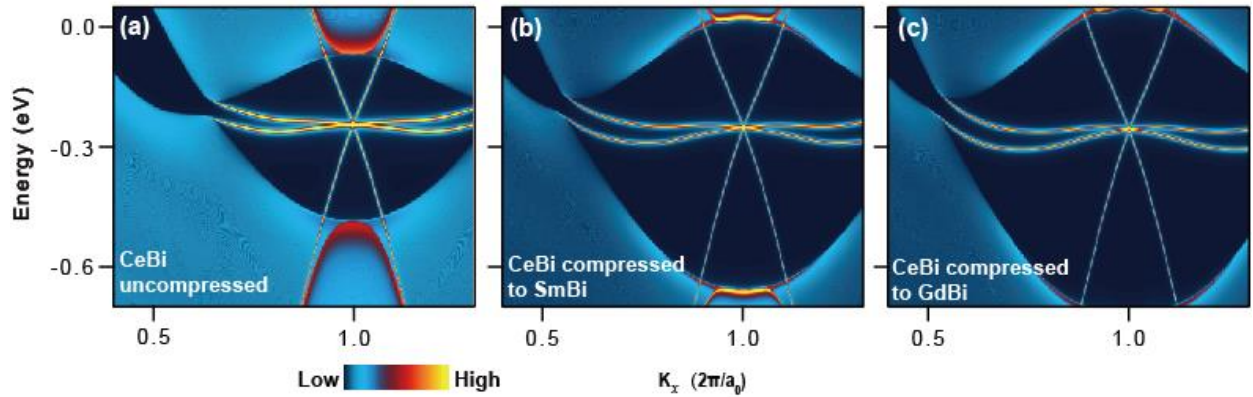

Fig. 7 (Color Online). Theoretical band structure of CeBi with various lattice constant, matched to those of bulk CeBi (a), SmBi (b) and GdBi (c), mimicking the pure pressure effect for CeBi. We used the experimental lattice constants ($a_0$) of 6.505, 6.35 and 6.311 Å for CeBi, SmBi and GdBi, respectively. The calculations were performed using the MBJ GF method.